%% file: workshop.tex
\documentclass{article}
\usepackage{spconf,amsmath,graphicx,hyperref}
\usepackage{amssymb,amsfonts}
\usepackage{textcomp}
\usepackage{xcolor}
\usepackage{booktabs}
\input{ju_math}
\usepackage{tikz}
\usepackage[numbers,sort&compress]{natbib}
\usepackage{hyperref}
\usepackage{subcaption}
\usepackage{multirow}
\usepackage{adjustbox} 
\usepackage{array}     
\usepackage{rotating}  
\usepackage{balance}
\usepackage{pdfx}

\title{Revolutionizing Diffusion MRI Microstructure Mapping via Global Inversion
}

\name{Yuxiang Wan$^1$, Hamza Farooq$^2$, Wenjie Zhang$^1$, Qiaozhi Huang$^1$, Lingjie Su$^1$, Christophe Lenglet$^2$, Ju Sun$^1$\thanks{Partially funded by NSF 2548082 and 2435911, NIH R01CA287413.}}
\address{$^1$Department of Computer Science and Engineering, University of Minnesota, Minneapolis, USA\\
$^2$Center for Magnetic Resonance Research, University of Minnesota, Minneapolis, USA\\
Emails: \{wan01530, faroo014, zhan7867, huan3189, su000342, clenglet, jusun\}@umn.edu}
\usepackage{IEEEtrantools}

\begin{document}
\bstctlcite{BSTcontrol}
%
\maketitle
\begin{abstract}
\input{sec/abstract}    
\end{abstract}
\input{sec/introduction}
\input{sec/related_work}

\input{sec/method}

\input{sec/experiment}

\clearpage 
\bibliographystyle{IEEEtran}

\bibliography{references}

\end{document}

%% file: ju_math.tex
\usepackage{graphicx,amsfonts,amscd,amssymb,bm,url,color,latexsym,bbm,amsthm,amsmath}
\usepackage{physics}

\usepackage{algpseudocode}
\usepackage{amsmath, amssymb}
\usepackage{tikz}
\usepackage{xcolor}
\usepackage{graphicx}
\usepackage{subcaption}

\allowdisplaybreaks
\usepackage{hyperref}
\usepackage[capitalize,nameinlink]{cleveref}
\crefname{section}{Sec.}{Secs.}
\crefname{table}{Tab.}{Tabs.}
\crefname{figure}{Fig.}{Figs.}
\crefname{theorem}{Thm.}{Thms.}
\crefname{appendix}{App.}{Apps.}

\newcommand{\mb}{\boldsymbol}

\newcommand{\mc}{\mathcal}

\numberwithin{equation}{section}

%% file: sec/abstract.tex
Diffusion MRI microstructure mapping (MM) is conventionally solved voxel by voxel, ignoring the fact that tissue microstructure forms a spatially organized field. This isolation leaves each estimation problem ill-posed and nonconvex. We instead cast MM as a single global inverse problem, reconstructing the entire parameter field jointly from all measurements of a subject. An untrained neural representation supplies implicit spatial priors and eases the nonconvex optimization, requiring no training data, while coregistered T1-weighted anatomy contributes structural guidance that is freely available in standard protocols. On both synthetic and in-vivo data, our method compares favorably with established voxel-wise and learning-based baselines, suggesting global inversion is a promising alternative.


%% file: sec/introduction.tex
\section{Introduction}

Diffusion magnetic resonance imaging (dMRI) probes tissue microstructures through measurements of water diffusion---called \emph{microstructure mapping} (MM)~\cite{stejskal1965spin,jones2010diffusion}. Physically, diffusion measurements $\mb{y}$ and tissue microstructures $\mb{x}$ are related by the forward process $\mb{y} \approx f(\mb{x})$, where $f$ is the \emph{highly nonlinear} diffusion process, and the approximation sign $\approx$ accounts for possible modeling errors and measurement noise. So, MM generally is a \emph{highly nonlinear} inverse problem~\cite{theory_and_parameter_estimation}. 

Hoping for better tractability, current MM methods attempt to \emph{simplify} the forward model by: (1) \textbf{local isolation}, where they divide the entire tissue volume into \textbf{localized} voxels, and further assume \textbf{independence} of measurements and microstructures between voxels---even if neighboring ones; (2) \textbf{structural abstraction}, where the estimation targets become \textbf{statistics or features} (i.e., image-derived phenotypes) of voxel-level microstructures---that parametrize their voxel-level forward biophysical models~\cite{Zhang2012NODDI,Alexander2010ActiveAx,Jelescu2022NEXI}, rather than the microstructures themselves. Based on such ``simplified'' forward models, they perform independent voxel-wise local inversion. 

However, local isolation may actually \textbf{harden} the overall MM problem, as (1) theoretically, it ignores potential global structures of the tissue volume, such as spatial smoothness, symmetries (e.g., for the entire brain volume), and spatial recurrence, all strong priors to tame the possibly ill-posed MM problems that might not have stable, unique solutions~\cite{Jelescu2016Degeneracy,theory_and_parameter_estimation}; and (2) computationally, each ``simplified'' voxel-level forward model remains highly nonlinear~\cite{Zhang2012NODDI,Alexander2010ActiveAx,Jelescu2022NEXI}, and hence natural optimization formulations for local inversion are invariably highly nonconvex problems---which may benefit from joint optimization with neighboring voxels due to the shared forward model and spatially correlated measurements. 

Therefore, in this paper, we pioneer a \textbf{global inversion} approach to MM, where the microstructures of the entire tissue volume are estimated \textbf{jointly and directly} from the measurements of the entire volume. This contrasts sharply with existing MM methods, whether optimization-based or learning-based, that are predominantly based on local inversion or, at best, on semi-local inversion; see \cref{sec:MM-methods}. We focus on \textbf{dMRI for human brains}, where the vast majority of public dMRI data are located around~\cite{HCP_data,alfaro2018image}. We base our method on the \textbf{optimization-based approach} for inverse problems---rather than the data-hungry and computation-heavy learning-based---by solving the classic optimization formulation of the form 
\begin{equation} \label{eq:map-generic}
\min\nolimits_{\mb{x}}
\underbrace{\ell\!\left(\mb{y},f(\mb{x})\right)}_{\text{data-fidelity loss}} + \lambda \underbrace{\Omega(\mb x)}_{\text{prior-inspired regularizer}}. 
\end{equation}
For the global forward model $f$, we aggregate voxel-level biophysical forward models~\cite{Zhang2012NODDI,Alexander2010ActiveAx,Jelescu2022NEXI} popularly used in the current local-inversion-based MM methods---we leave the use of more faithful global models to future work.  

To solve MM well, we need to integrate suitable global priors into  \cref{eq:map-generic} and also handle the nonconvexity caused by the nonlinear $f$. Toward this,  we consider \textbf{untrained neural priors}~\cite{alkhouri2025understanding} for their dual power to provide implicit global priors and to empower global optimization despite nonconvexity. Moreover, we add a structural prior from the T1-weighted MRI, which often accompanies dMRI measurements and reveals related microstructures~\cite{slator2021combined}. Together, we contribute the first optimization-based global-inversion MM method, opening a new chapter for dMRI research.

%% file: sec/related_work.tex
\section{Related Work and Background}

\subsection{Untrained Neural Priors (UNPs)}

UNPs reparametrize the estimation target $\mb x$ in \cref{eq:map-generic} as a function of learnable deep neural networks (DNNs). For example, in deep image prior (DIP), $\mb x = G_{\mb \theta}(\mb z)$ where $G_{\mb \theta}$ is a learnable DNN and $\mb z$ is a frozen seed~\cite{dip}; in implicit neural representation (INR), $\mb x = \mc D \circ G_{\mb \theta}$, where $G_{\mb \theta}$ is a coordinate-input DNN representing the continuous version of $\mb x$ and $\mc D$ is a discretization operator~\cite{siren,tancik2020fourier,Saragadam2023WIRE}. These $G_{\mb \theta}$'s are typically heavily overparametrized, making global optimization more likely---motivated by optimization theories for training overparameterized DNNs~\cite{berner2021modern}. Moreover, specific architecture choices in $G_{\mb \theta}$ together with gradient-based optimization dynamics often favor ``simple'' solutions, e.g., those with spatial smoothness or strong periodicity~\cite{dip,alkhouri2025understanding}. The double benefits of UNPs in global optimization and implicit structural biases have recently enabled numerous successes in the solution of inverse problems, particularly difficult scientific ones~\cite{DeepRandomProjector,Wang2023EarlyStopping,li2021self,Tayal2021PhaseRetrieval,Zhuang2023DoubleDIP,blinddeblur}. In addition, compared with these data-driven priors that need massive training datasets for pretraining, e.g., pretrained generative priors~\cite{wang2024dmplug,wan2025fmplug,wan2026fmplug,wang2026temporal}, UNPs are lightweight and data-free. Due to the performance and data advantages of UNPs, we choose UNPs for our global MM method. 

\subsection{Microstructure Mapping (MM) for dMRI}
\label{sec:MM-methods}

\textbf{Optimization-based approach.} 
All popular MM methods in this family are based on local inversion, and focus on developing meta-heuristic global optimization algorithms and software framework to solve the voxel-level nonconvex problems for high-quality solutions. For example, MIX combines variable elimination and genetic-algorithm-based global search to secure a good initialization \cite{MIX}; AMICO integrates brutal-force search with convex reformulation~\cite{AMICO}; DMIPY and cuDIMOT implement different biophysical models and their corresponding MM methods in unified software frameworks with GPU acceleration~\cite{Fick2019Dmipy,gpu_accelerate}. Despite their popularity, they never exploit any spatial priors between voxels, a salient feature in our global inversion approach.

\vspace{0.5em}
\noindent \textbf{Learning-based approach.} 
Most supervised MM methods \cite{Golkov2016qDL,Ye2017MEDN, Zheng2022METSC,problem_population_based} focus on local inversion and learn the inverse mapping from voxel-level measurements to model parameters directly with paired (measurements, parameters) datasets. They differ mostly in their model design, ranging from multi-layer perceptrons to algorithm-informed unrolling and cascades. Recent ``self-supervised'' methods bypass paired datasets: DIMOND~\cite{DIMOND} form their learning objectives based on cycle consistency at the voxel-level (measurements $\to$ parameters {\footnotesize $\stackrel{f}{\to}$} measurements) and takes both the current voxel's and neighboring voxels' measurements, explicitly modeling local spatial correlation; \cite{Hendriks2025INR} uses INR to represent the whole brain's microstructures but focuses only on the standard model for white matter---the sole existing global inversion method as far as we know. 

We note that \textbf{\cite{DIMOND,Hendriks2025INR} can also be treated as optimization-based methods on the entire tissue volume}, the exact point of view that we take in this paper for global inversion. \cite{Hendriks2025INR} is close in spirit to our method, although they focus on a different biophysical model than ours (standard model vs our NODDI). Moreover, we also enhance our INR with wavelet-inspired modification to better recover high-frequency details and T1-weighted MRI structural guidance.

%% file: sec/method.tex
\section{Our Global Inversion Method}

Our goal is to estimate the microstructures of the entire brain together from all dMRI measurements, i.e., MM based on global inversion. Our estimation is based on the optimization formulation in \cref{eq:map-generic}: we describe the forward model in \cref{sec:fwd-model}, the IND-based formulation in \cref{sec:formulation}, the details of the IND model in \cref{sec:model-details}, and the T1-weighted MRI regularization in \cref{sec:anatomy-modulation}. 

\subsection{NODDI Forward Model}
\label{sec:fwd-model}

Although we treat all brain microstructures as a single object, we form our global forward model $f$ by aggregating the NODDI model at the voxel-level~\cite{Zhang2012NODDI}, i.e. the ``global'' NODDI model effectively takes localized, voxel-level measurements, and the ``global'' measurements are an aggregation of all such voxel-level measurements. Specifically, given location $\mb r$, gradient direction $\mb g_i$ and gradient strength $b_i$, 
\begin{equation}
    y_i(\mb r)
    \doteq 
    {S(\mb r,b_i,\mb g_i)}/{S_0(\mb r)}
    \label{eq:normalized-signal}
\end{equation}
is the normalized measurement: $S(\mb r,b_i,\mb g_i)$ is the direct measurement, and $S_0(\mb r)$ is the non-diffusion-weighted signal. NODDI models this signal as a mixture of intra-cellular, extra-cellular, and isotropic compartments~\cite{Zhang2012NODDI}:
\begin{align}
y_i(\mb r) & \approx F_{\mathrm N}(\mb\theta(\mb r);b_i,\mb g_i)  \\
\doteq & 
 (1-f_{\mathrm{iso}}) \left[ f_{\mathrm{ic}}A_{\mathrm{ic}} + (1-f_{\mathrm{ic}})A_{\mathrm{ec}} \right]  + f_{\mathrm{iso}}A_{\mathrm{iso}}. 
\label{eq:noddi-forward}
\end{align}
Here, $\mb\theta(\mb r) \doteq \{f_{\mathrm{ic}}(\mb r),f_{\mathrm{iso}}(\mb r), \kappa(\mb r),\mb\mu(\mb r)\}$, where $f_{\mathrm{ic}}(\mb r)$ and $f_{\mathrm{iso}}(\mb r)$ are the intra-cellular and isotropic fractions, respectively, $\kappa$ is the Watson concentration, and $\mb\mu$ is the mean neurite orientation. $A_{\mathrm{ic}}, A_{\mathrm{ec}}, A_{\mathrm{iso}}$ denote the signal attenuation of the intra-cellular, extra-cellular, and isotropic compartments, respectively, each a nonlinear function of $(b_i,\mb g_i)$ and $\mb\theta(\mb r)$. 

\subsection{Our Global-Inversion Formulation}
\label{sec:formulation}

Our INR model takes any normalized location coordinate $\mb \gamma(\mb r)$ and predicts the NODDI parameters $\{f_{\mathrm{ic}}(\mb r), \allowbreak f_{\mathrm{iso}}(\mb r), \allowbreak \kappa(\mb r), \allowbreak \mb\mu(\mb r)\}$ with the constraints $0\leq f_{\mathrm{ic}},f_{\mathrm{iso}}\leq1$, $\kappa\geq0$, and $\|\mb\mu\|_2=1$. We write it as 
\begin{equation}
    \mb\theta_\psi(\mb r)
    \doteq 
    \Pi \circ 
    G_\psi\!\left(\mb \gamma(\mb r)\right),
    \label{eq:siren-field}
\end{equation}
where $G_\psi$ is the trainable network (detailed in \cref{sec:model-details} below) with four prediction heads and $\Pi$ denotes the final activations to ensure that the predicted parameters meet their respective constraints. We use the $\mathrm{sigmoid}$ function for $f_{\mathrm{ic}}(\mb r), \allowbreak f_{\mathrm{iso}}(\mb r)$, $\cot(\pi/2 \cdot \mathrm{sigmoid}(\cdot))$ function for $\kappa$, and $\ell_2$ normalization for $\mb \mu$. Moreover, let $\Omega$ denote the set of voxel locations, and $\mc I$ the index set of measurement gradient vectors. Then our INR-based global inversion formulation is 
\begin{align}
&\min\nolimits_\psi \tfrac{1}{|\Omega|\,|\mc I|} \sum_{\mb r\in\Omega} \sum_{i\in\mc I}
\ell\!\left( y_i(\mb r), F_{\mathrm N}\!\left( \mb\theta_\psi(\mb r);b_i,\mb g_i \right)\right).
\label{eq:test-time-objective}
\end{align}
Our formulation in \cref{eq:test-time-objective} is very similar to that in \cite{Hendriks2025INR}, with two main differences: (1) Motivation-wise, they come from joint local inversion with the INR promoting spatial correlation, whereas ours come from global inversion---although with a compromised global forward model. We argue that our global inversion framework represents a paradigm shift for dMRI MM: with more faithful, non-locally-isolated global forward models, global inversion may demonstrate substantial performance advantages; and (2) Forward-model-wise, we focus on the NODDI model, vs. the standard model in \cite{Hendriks2025INR}. Moreover, we adopt a wavelet-inspired INR model that better encodes high-frequency components (\cref{sec:model-details}), and integrate an MRI-informed regularization on the global structure (\cref{sec:anatomy-modulation}). 


\subsection{The Details of Our INR Model}
\label{sec:model-details}

We adapt the Wavelet INR (WIRE) model~\cite{Saragadam2023WIRE} to form our INR model $G_\psi$. WIRE model is a complex-valued multi-layer perceptron (MLP) with the complex Gabor activation 
\begin{align}
    \sigma(z; \omega, s) \doteq e^{\mathrm{i} \omega z - \abs{sz}^2},
    \label{eq:gabor_act}
\end{align}
where $\omega$ and $s$ control the primary frequency and magnitude $\sigma(z)$ can represent. The real part of the final output is fed into $\Pi$ in \cref{eq:siren-field}. To allow $G_\psi$ to adapt to the localized frequency and magnitude scales, we make the $\sigma$ and $s$ of each neuron learnable---jointly optimized with the standard MLP weights through \cref{eq:test-time-objective}  and  initialized with WIRE's default setting $\omega=30$ and $s=10$. 
We call this model
\emph{Ada-WIRE}. 

\subsection{MRI-Based Regularization}
\label{sec:anatomy-modulation}

T1-weighted MRI delineates tissue boundaries via longitudinal relaxation
contrast, at higher resolution than dMRI and largely independently of the
diffusion signal. Because T1-weighted scans are routinely acquired
alongside dMRI in standard protocols, this structure information is
typically available at no additional acquisition cost~\cite{slator2021combined,glasser2013hcp}. 

To encode the T1-informed structure information, we develop a separate modulation model that interacts with the WIRE model. To construct the modulation model, we first resample the T1-weighted MRI with the same spatial resolution as that of dMRI, so that we can index them with the same spatial coordinate system. Writing the resampled MRI as $T$, we define a learnable voxel-level latent vector as 
\begin{multline}
  \mb h_{\mb \gamma}(\omega_{\mb \gamma}, W_{\mb \gamma}, b_{\mb \gamma}) \\
  = \sin\!\big( \omega_{\mb \gamma} \left( W_{\mb \gamma} [\, \mb \gamma; T(\mb \gamma); \; \|\nabla T (\mb \gamma)\|  \,] + b_{\mb \gamma} \right) \big),
  \label{eq:anatomy-latent}
\end{multline}
which takes the coordinate vector $\mb \gamma$, MRI intensity $T(\mb \gamma)$, and MRI gradient magnitude $\|\nabla T (\mb \gamma)\|$ as inputs. The gradient magnitude is to emphasize tissue boundaries, which are likely the most critical information we can obtain from $T$. Next, on top of these latent vectors, we define layer-wise modulation vectors as  
\begin{align}
  \alpha_\ell(\mb \gamma) 
  & = 1 + s_\alpha(\mb \gamma) \tanh\!\left( W^\alpha_\ell (\mb \gamma) h_{\mb \gamma} + b^\alpha_\ell (\mb \gamma) \right),
  \\
  \beta_\ell (\mb \gamma)
  & = s_\beta (\mb \gamma) \tanh\! ( W^\beta_\ell (\mb \gamma) h_{\mb  \gamma} + b^\beta_\ell (\mb \gamma)),
  \label{eq:film-params}
\end{align}
inspired by feature-wise linear modulation (FiLM)~\cite{FiLM}. The final modulation takes the form 
\begin{equation}
  \widetilde z_\ell (\mb \gamma) = \alpha_\ell (\mb \gamma) \odot z_\ell (\mb \gamma) + \beta_\ell (\mb \gamma),
  \label{eq:film-preact}
\end{equation}
where $z_\ell$ denotes the pre-activation outout of the WIRE model at the $\ell$-th layer. We now arrive at the modulated WIRE model, which we call \emph{Ada-WIRE Anatomy}. 

We set $\omega_{\mb \gamma} = 3$ and $s_\alpha = s_\beta = 0.1$, and initialize all FiLM weights and biases to zero, so $\alpha_\ell = 1$ and $\beta_\ell = 0$ at initialization: optimization begins from the unmodulated WIRE, and the bounded $\tanh$ function confines the modulation strength to roughly $\pm 10\%$, so the MRI structures bias the solution without overriding the data consistency. The final linear readout is left unmodulated.

%% file: sec/experiment.tex
\section{Experiment}

\begin{table}[!htbp]
  \centering
  \caption{MM performance on the simulated dataset. Lower is better; bold indicates the best mean. ODI denotes the orientation-dispersion index. }
  \label{tab:simulated-mse-full}
    \resizebox{\columnwidth}{!}{%
    \begin{tabular}{lccc}
      \toprule
      Method & ODI MSRE $\downarrow$
        & $v_{\mathrm{ic}}$ MSRE $\downarrow$
        & $v_{\mathrm{iso}}$ MSRE $\downarrow$\\
      \midrule
      AMICO--NODDI
        & $.380 \pm .120$ & $.015 \pm .002$ & $.908 \pm .146$ \\
      MIX
        & $.044 \pm .018$ & $.010 \pm .003$ & $1.077 \pm .210$\\
      DIMOND
        & $.186 \pm .197$ & $\mathbf{.004 \pm .002}$ & $\mathbf{.117 \pm .078}$\\
      \midrule
      WIRE
        & $.025\pm .013$ & $.010 \pm .005$ & $.868 \pm .382$\\
      Ada-WIRE (Ours)
        & $\mathbf{.022\pm .014}$ & $.009 \pm .004$ & $.870 \pm .372$\\
      \bottomrule
    \end{tabular}
  }
  \vspace{-.5em}
\end{table}

\begin{table}[!htbp]
  \centering
  \caption{Normalized diffusion-signal reconstruction error on the HCP dataset. Values are subject-level mean $\pm$ sample standard deviation; lower is better.}
  \label{tab:hcp-signal-reconstruction}
  \setlength{\tabcolsep}{4pt}
   \resizebox{0.8\columnwidth}{!}{%
  \begin{tabular}{@{}lcc@{}}
    \toprule
    Method  & MAE $\downarrow$ & MSRE $\downarrow$ \\
    \midrule
    AMICO--NODDI & $.061 \pm .002$ & $.141 \pm .028$ \\
    MIX & $.069 \pm .006$ & $.136 \pm .008$ \\
    DIMOND & $.066 \pm .006$ & $.115 \pm .015$ \\
    \midrule
    WIRE (Ours) & $.058 \pm .006$ & $.123 \pm .010$ \\
    Ada-WIRE (Ours) & $.055 \pm .002$ & $.122 \pm .009$ \\
    Ada-WIRE, Layer-wise (Ours) & $\mathbf{.053 \pm .003}$ & $\mathbf{.110 \pm .010}$ \\
    \bottomrule
  \end{tabular}
  }
  \vspace{-1em}
\end{table}

\subsection{Setup}

\textbf{Data.}   
\textbf{(1) Synthetic dataset}: 
We generate $20$ synthetic volumes following the NODDI model, with $2, 6$ and $10$ spatially distinct regions, respectively. Each synthetic volume consists of $10,000$ voxels ($22\times22\times21$, 1.25~mm isotropic grid)  with fiber orientation $(\theta,\phi)\in[.01,\pi]$, ODI$\;\in[.02,.8]$, and $v_{\mathrm{ic}}\in[.2,.8]$ varying smoothly within each region ($v_{\mathrm{iso}}=.01$, $d_{\parallel}=1.7$, $d_\perp=d_\parallel(1-v_{\mathrm{ic}})$, isotropic diffusivity $3.0~\mu\mathrm{m}^2\mathrm{ms}^{-1}$). Each voxel has $81$ simulated measurements ($9$ at $b=0$, $24$ at $b \approx700$, $48$ at $b \approx 2000~\mathrm{s\,mm}^{-2}$). We simulate both noise-free and noisy measurements, where the latter contains Rician noise at SNR~20. We do not have T1-weighted MRI for this data. 
\textbf{(2) the Human Connectome Project (HCP) dataset}: We take data from $10$ subjects of the WU--Minn HCP Young Adult dataset. It contains $288$ measurements per voxel ($18$ at $b{=}0$, and $270$ diffusion-weighted split among $b=1000,2000,3000~\mathrm{s\,mm}^{-2}$, $90$ directions/shell) at $1.25$~mm isotropic resolution~\cite{HCP_data}. Each subject comes with an accompanying T1-weighted MRI scan. 

We note that the number of subjects may seem small in our datasets, but it is comparable to standard practice in the dMRI literature---as their evaluation focuses on voxel-level performance, and datasets of such sizes already contain a huge number of 3D voxels. In future work, we will substantially scale up the number of subjects.


\begin{figure}[h]
  \centering
  \includegraphics[width=0.95\columnwidth]{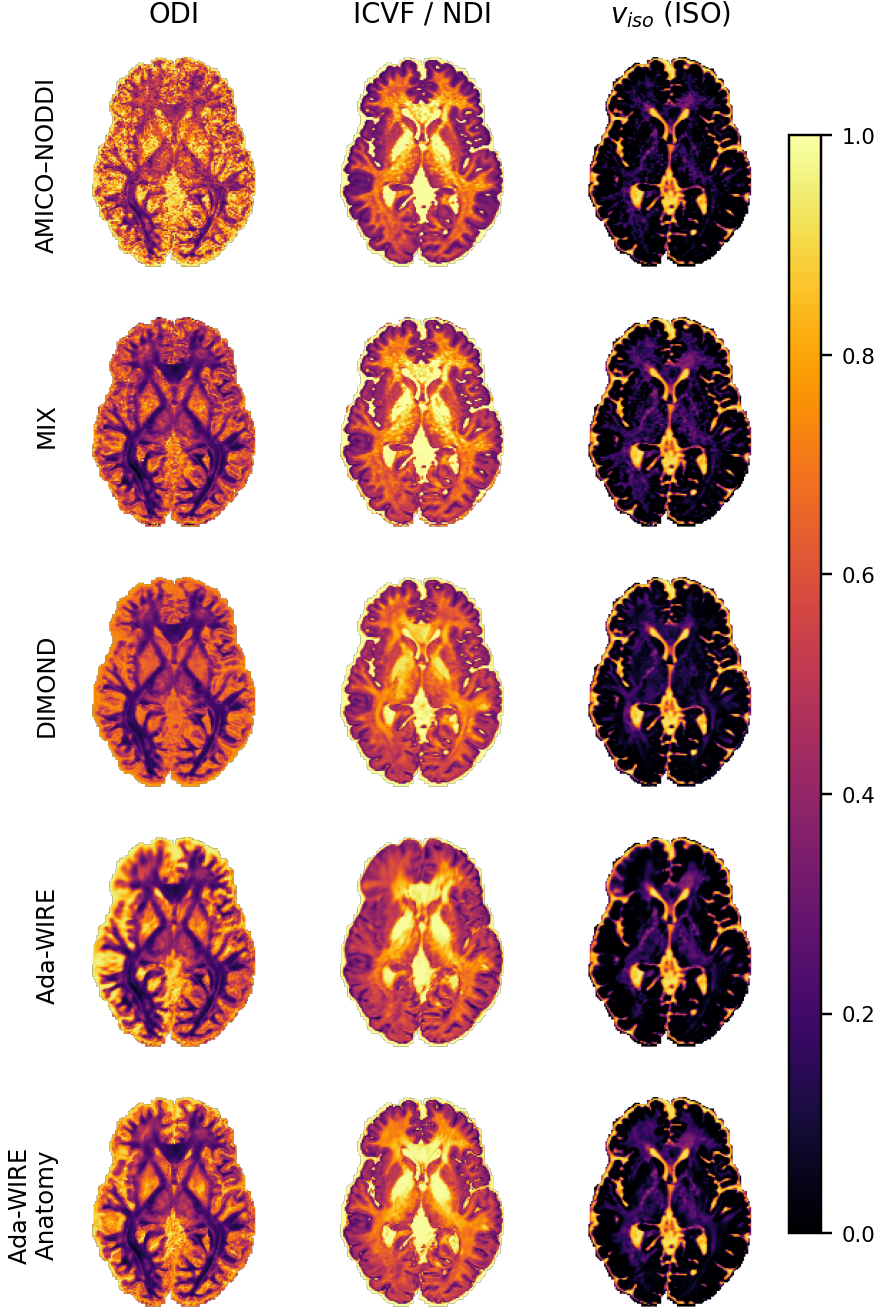}
  \caption{NODDI microstructure maps for the HCP subject 100408, axial slice $k=60$. Rows: AMICO--NODDI, MIX, DIMOND, Ada-WIRE, and anatomy-modulated Ada-WIRE. Columns: ODI, intra-neurite fraction (ICVF/NDI), and isotropic fraction ($v_{\mathrm{iso}}$). All panels share a $[0,1]$ scale.}
  \label{fig:hcp-subject-visualization}
\end{figure}

\vspace{1em}
\noindent \textbf{Baseline methods \& evaluation metrics.} 
Baseline methods include voxel-wise AMICO--NODDI \cite{AMICO}, MIX \cite{MIX}, and per-subject convolutional DIMOND \cite{DIMOND}. For our methods, we include three variants: WIRE~\cite{Saragadam2023WIRE}, Ada-WIRE ( \cref{sec:model-details}), and Ada-WIRE Anatomy (\cref{sec:anatomy-modulation}). We adopt standard evaluation metrics in the dMRI literature \cite{AMICO, farooq2026diffusion}: three variants of the voxel-level mean square relative error (MSRE) between the estimated parameters and the groundtruth parameters for the synthetic dataset; MSRE and mean absolute error (MAE) between the physical measurements and the estimated measurements, i.e., assessing the data-fitting quality, as no groundtruth parameters available. Our HCP analysis emphasizes visual spatial organization.

\subsection{Results}
\cref{tab:simulated-mse-full} reports our results over $20$ volumes of the synthetic datasets. Ada-WIRE achieves the lowest ODI MSRE and improves all three parameter errors over AMICO--NODDI and MIX. DIMOND achieves the lowest errors for $v_{\mathrm{ic}}$ and $v_{\mathrm{iso}}$, but its larger ODI standard deviation indicates less consistent recovery across samples. 

On the HCP dataset, we emphasize visual spatial organization because in-vivo parameter ground truth is unavailable. \Cref{fig:hcp-subject-visualization} shows the comparison for HCP subject 100408. In the ODI maps, Ada-WIRE Anatomy delineates low-ODI tract-like bands with clearer boundaries and a more smooth spatial extent. AMICO--NODDI and MIX exhibit speckled variations that interrupt or obscure these patterns. DIMOND produces smoother maps, but some narrow bands are less distinctly separated from the surrounding tissue. Ada-WIRE also preserves coherent bands, while the MRI-based anatomical modulation makes selected boundaries and fine structures more distinct. \Cref{tab:hcp-signal-reconstruction} provides a secondary assessment of the measurement consistency. Ada-WIRE Anatomy achieves the lowest mean MAE and MSRE, indicating that the visually coherent maps also fit the acquired diffusion signal well.  